# Performance benchmarking of an ultra-low vibration laboratory to host a commercial millikelvin scanning tunnelling microscope

*Yande Que,[1] Amit Kumar,[1] Michael S. Lodge,[1] Zhengjue Tong,[1] Marcus Lai Kar Fai,[1] Wei Tao,[1] Zhenhao Cui,[1] Ranjith Shivajirao,[1] Junxiang Jia,[1] Siew Eang Lee,[2] and Bent Weber[1*]*

[1]Division of Physics and Applied Physics, School of Physical and Mathematical Sciences, Nanyang Technological University, Singapore 637371

[2]Department of the Built Environment, College of Design and Environment, National University of Singapore, 4 Architecture Drive, 117566 Singapore

E-mail: b.weber@ntu.edu.sg

Ultra-low temperature scanning tunnelling microscopy and spectroscopy (STM/STS) achieved by dilution refrigeration can provide unrivalled insight into the local electronic structure of quantum materials and atomic-scale quantum systems. Effective isolation from mechanical vibration and acoustic noise is critical in order to achieve ultimate spatial and energy resolution. Here, we report on the design and performance of an ultra-low vibration (ULV) laboratory hosting a customized but otherwise commercially available 40-mK STM. The design of the vibration isolation consists of a T-shaped concrete mass block (~55t), suspended by actively controlled pneumatic springs, and placed on a foundation separated from the surrounding building in a "room-within-a-room" design. Vibration levels achieved are meeting the VC-M vibration standard at >3 Hz, reached only in a limited number of laboratories worldwide. Measurement of the STM's junction noise confirms effective vibration isolation on par with custom built STMs in ULV laboratories. In this tailored low-vibration environment, the STM achieves an energy resolution of 43 μeV (144 mK), promising for the investigation and control of quantum matter at atomic length scales.

# 1. Introduction

Isolation from acoustic and vibrational noise [1] plays an essential role in modern ultra-high precision instrumentation and measurement, given their deleterious effects on spatial and energy resolution. Scanning tunnelling microscopes (STM) with their unrivalled capability in revealing structural and electronic information at atomic length scales [2–5] are particularly susceptible to disturbances as the tunnelling current is exponentially sensitive to the tip-sample separation (typically <1 nm). For ultimate performance, sophisticated damping stages are therefore needed to effectively isolate from mechanical noise and vibrations. Especially dilution refrigerator (DR) cooled STMs require an additional pumping system to cool down the system by liquifying and circulating the $^3$He/$^4$He mixture. Such pumping system generates significant noise and vibrations during operation. Hence, additional vibration isolation of the STM from its pumping system is required for optimized performance.

Very recently, DR-cooled mK-STMs have become commercially available for the first time, with only a handful (<10 at the time of writing) commissioned worldwide. Here, we benchmark the performance of such a system - a UNISOKU USM1600 mK-STM - hosted within a tailored ultra-low vibration (ULV) laboratory space. Our USM1600 system is customized with a two-axis vectorized superconducting magnet (9T in vertical direction, and 4 T in horizontal direction, with 3.5 T in full vector rotation) and provides cooling to a base temperature of 40 mK at the STM head. The wires connecting to sample and STM tip are RF coaxial cables, allowing signal transmission up to 10 GHz frequencies. There are a total of six wires connecting to the STM sample stage, customized with six electrodes, allowing transport measurement for nanoelectronic devices in addition to standard STM operation. To employ a capacitance-mediated tip approach onto micron-sized samples, a customized tip holder with a grounded shield is integrated with the STM. A position sensor is integrated into STM head, allowing to record the vertical position of the STM tip with respect to the sample for easier tip approach, especially onto small samples.

## 2. Laboratory design

Since the invention of the STM [6–8], various designs and damping stages have been employed to achieve ultra-low vibration levels [9–11]. A compact and rigid design of STM heads combined with springs and eddy current damping have been widely used in both cryogenic and non-cryogenic STM without magnetic coils. Heavy steel tables or frames, resting on pneumatic legs are commonly used as a passive damping stage, in order to maximize STM performance in noisy laboratory environments, and/or if eddy current damping cannot be employed due to the presence of a magnet. To further isolate from ground or structure borne vibrations arising in the laboratory's vicinity, massive foundations [10,11] have been employed ideally hosted in an isolated building. However, often separate builidings are not practical or financially not feasible, so foundations, that are structurally isolated from an existing building have been considered. Further passive [9,12–14] or even active [15,16] isolation of heavy masses or pneumatic spring systems have been considered in much fewer laboratories, to achieve ultimate performance. In general, different damping stages are usually combined to achieve the possible lowest vibration levels. These include one or more layers of massive inertia block(s), supported by pneumatic isolators and/or active dampers [12–19].

**Figure 1** shows a CAD schematic of our ULV laboratory, located on the ground floor of a five-story building on the Nanyang Technological University (NTU) campus in Singapore. The greater lab space features separate rooms for the lab's dedicated mechanical and electrical services (M&E), a liquid helium (LHe) recovery and liquefaction plant, and a separate room for measurement control (**Figure 1b**). For the ULV lab itself, a "room-within-a-room" solution was adopted, that is structurally decoupled from the surrounding building. To achieve this, an area of $(7.8 \times 8.6)$ m$^2$ was excavated to 4.0 m depth in the center between four of the building's

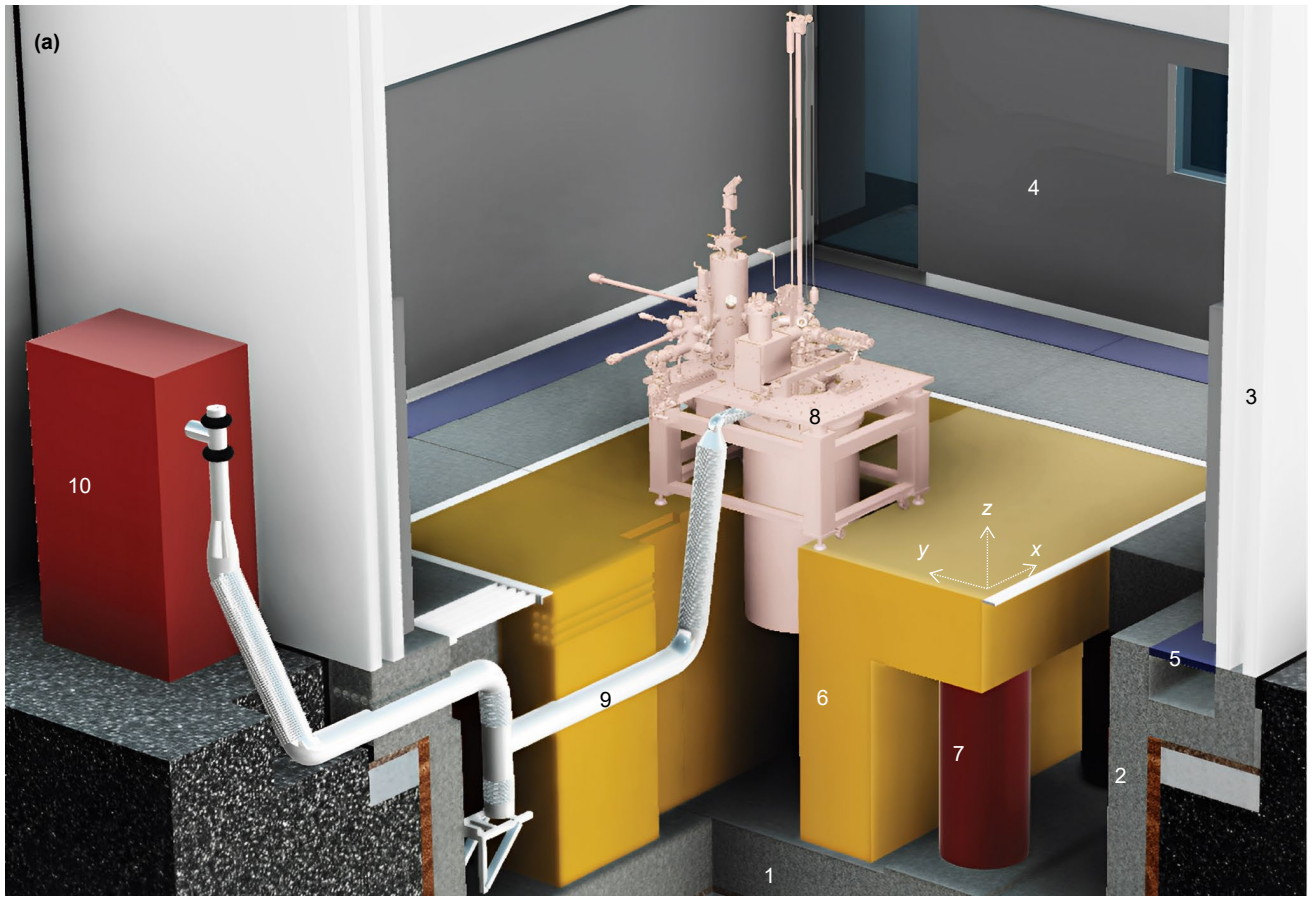

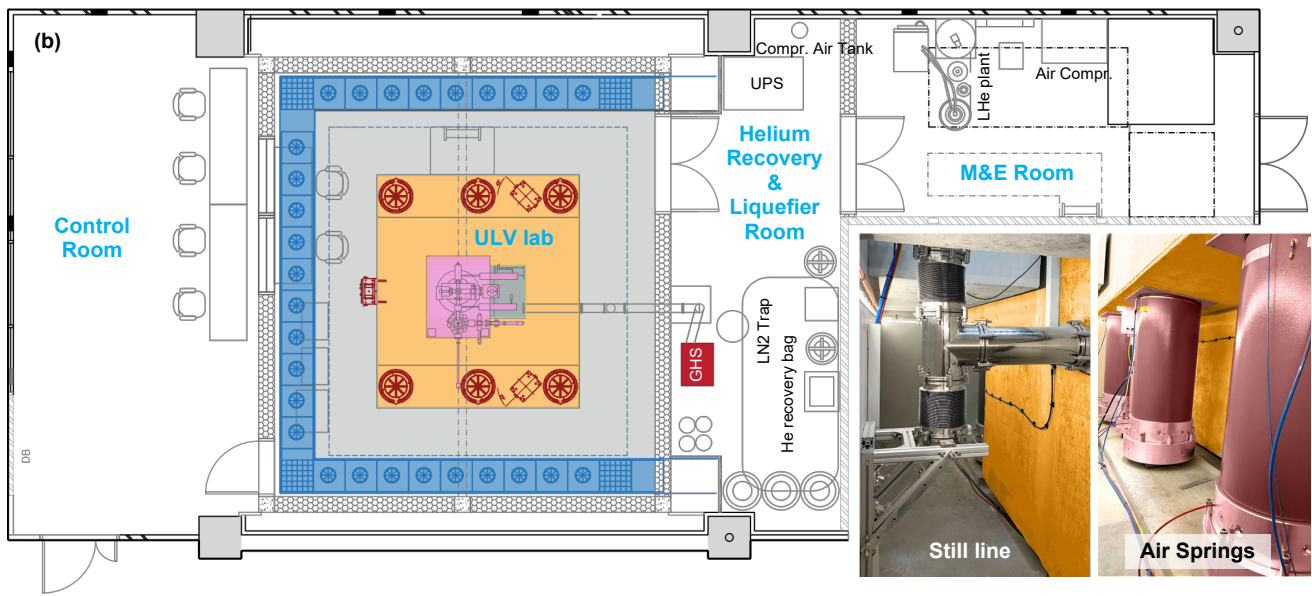

**Figure 1. An ultra-low vibration laboratory for mK-STM at NTU Singapore. (a)** CAD schematic of the lab. (1,2) Concrete foundation base plate (1.5 m thick) and cradle (2) structurally isolated from the surrounding building, (3) Double-layer brick walls separated by an air gap, (4) Acoustic panels. (5) Air-conditioning supply points, (6) T-shaped mass block (~55t), (7) Active pneumatic isolators with low resonance frequency, (8) DR-cooled STM, (9)

DR still line, (10) Gas handling system (GHS). The coordinate system used for the vibration measurement is indicated. **(b)** Floor plan of the laboratory consisting of ULV laboratory, control room, helium recovery and liquefier room, and M&E room. The inserts show photos of the T-junction isolator for the DR still line and the air spring isolators, respectively.

structural support pillars (gray squares in **Figure 1b**). A 1.5 m thick concrete foundation base plate (1) was cast onto a layer of 250mm/50mm compacted rubble placed on the bare soil, forming a raft foundation upon which the entire laboratory room was built. Next, a reinforced concrete cradle was erected on this foundation, hosting a T-shaped mass block (6) with a weight of ~55t. To avoid forces on the superconducting magnet, we have decided for non-magnetic fiber glass reinforcement of the concrete block instead of conventional steel rebar. Double-layer brick walls (3) with air gaps for acoustic isolation were erected on the foundation cradle surrounding the block, stabilized by vertical steel columns, and covered by a concrete roof at a height of 5.9 m. Acoustic panels were used to cover the lower half of the ULV lab walls up to 2.5 m in height to void reverberation and standing sound waves. A fully integrated envelope, including asoustic door for connecting adjacent spaces, was designed to provide a sound insulation value of STC 55. The measured ambient noise levels in the ULV lab after completion were found to meet the NC20 acoustic standard [20] with air-conditioning system operating under normal conditions. The low internal ambient noise level eliminates acoustic waves excitation of the internal structures and equipment.

The T-shaped concrete mass block is supported by six actively controlled pneumatic springs (7). We have chosen active pneumatic isolators over piezoelectric isolation systems as the latter require additional passive isolators to absorb the static mass due to the low actuator force. This can be effective at low frequencies but can be found to transmit noise at higher frequencies. Active pneumatic isolators [19] have a soft spring combined with active vibration control, allowing for the isolation of large load weights over a wide frequency range. The active isolators used (Air-springs BiAir4-OC-1, and horizontal elements HAB-95000 by Bilz,

Germany) have been customized to a low resonance frequency of 1.2 Hz in vertical direction, and 0.7 Hz in both horizontal directions. The DR-cooled mK-STM system (8) itself is mounted on a stainless-steel table with pneumatically damped legs that have been designed to have a resonance frequency between 3~4 Hz, considerably above the resonance frequency of the active air spring isolation of the block. The STM head of the USM1600 is hung from three soft springs, with typical resonance frequencies around 5~6 Hz, above those of both table and block. Continuous operation of the STM's dilution cryostat at base temperature requires a closed-loop gas handling system (GHS) (10) that continuously circulates He gas. The associated pumps include one oil-free rotary pump for the evacuation of the 1K-pot to condense the $^3$He/$^4$He mixture, one turbo molecular pump and, one oil-free rotary pump for the circulation of the mixture. These pumps generate acoustic and vibration impact, which might transmit to the STM through the air, ground, or circulation pipes. To minimize this impact, we have placed the GHS including pumps within an adjacent space [14] next to the ULV lab (see **Figure 1b**). All connecting lines between two spaces have been anchored into the concrete mass of cradle and isolation block, respectively, as shown in **Figure 1a**. They have been further isolated from the gas handling system and equipment, respectively, by long formed bellows. The DR still line (9) (diameter: 160 mm) connecting the circulation pumps and the still of the dilution cryostat was isolated by a T-junction with two counteracting soft bellows to allow the isolation block to move freely in all directions when suspended. All other pumping lines were isolated by long formed bellows.

Conventional top-down air-conditioning systems (ACS), common to many experimental labs, generate acoustic noise or turbulent airflow that can have deleterious effect on a softly damped suspended slab, or may exert air pressure directly onto experimental platforms. Shutting down the ACS during measurement is not always practical as it may induce a thermal gradient in the lab space over time, and give rise to increased humidity levels. To mitigate this problem, we have employed an upward flow ACS with air circulation and autonomous control separate from the building's central air conditioning system. In the ULV lab, supply air at a volume flow rate of 28 l/s (100 m$^3$/h) is injected from 27 low-velocity outlet points, cast at floor level and at a lateral distance of ~1 m away from the floating slab, avoiding the use of air ducts that generate duct-borne noise and vibratoin. The final air velocity (degree of turbulence) were measured to be 0.31 m/s (27%) at a position 0.5 m away from the outlet in lateral distance and 50 mm above the floor, allowing an efficient air-circulation to cool the ULV laboratory's volume to 23 °C. The return air is collected in a large volume above a false ceiling (0.9 m below the concrete ceiling of the lab) via nine (0.6 × 0.6) m$^2$ grid panels and returned via a single vibration-isolated exhaust duct above. Negative pressure within the false ceiling space ensures vertical airflow and minimizes air turbulence. Autonomous local control further allows to throttle the airflow in the lab or switch it off entirely if needed for selected high-precision experiments. In addition to the achievement of a low noise environment, the displacement ventilation system can also enhance the energy efficiency of the laboratory.

## 3. Results and discussion

*3.1 Floor vibration levels*

To evaluate the floor vibration levels, we have carried out seismic measurements on both the concrete base plate (1) and the floating slab (6). All measurements were performed in peak-hold mode (except where specifically indicated), recorded over typical time windows of 1 min using Wilcoxon 731A accelerometers. For all the measurements presented in **Figure 2** and **3**, the air-conditioning airflow was switched off, with all other utilities running, reflecting the lowest vibration levels achievable for DR-cooled STM. Vibration impact by air-conditioning and air compressor will be analyzed in **Figure 4**.

In **Figure 2**, we benchmark the vibration levels of the NTU ULV laboratory against those published for state-of-the-art STM laboratories internationally. To be consistent with international vibration standards [17] and published literature [9], we present the RMS of the vertical floor velocity in 1/3-octave frequency bands [21]. We find that the vibration levels of the NTU ULV laboratory (floating slab) are well below those of the VC-M vibration standard above 3 Hz. Overall, vibration levels are among the lowest compared to few labs internationally that feature a floating block, and for which data could be found.

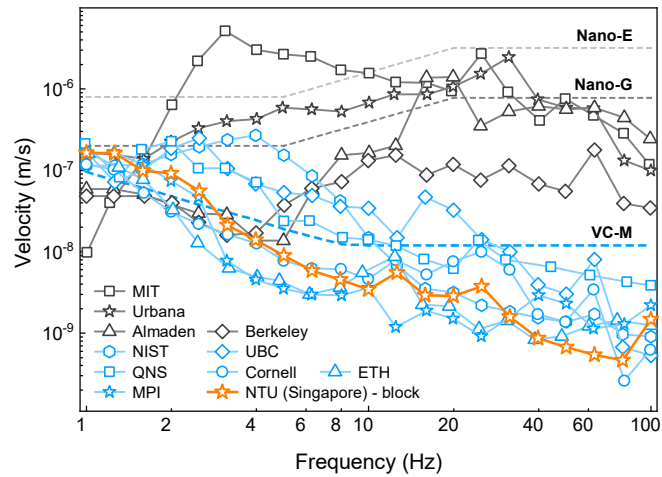

**Figure 2. Floor vibration levels of NTU's ULV laboratory.** Performance benchmark against STM laboratories internationally. Spectra of the root mean square (RMS) of the vertical ($Z$) floor velocity are plotted in one-third octave bands. Data plotted as black squares (MIT), triangles (Urbana-Illinois), circles (IBM-Almaden), diamonds (Berkeley) and blue circles (Cornell) are reproduced from Ref. [21]. Data shown as blue hexagons (NIST-Maryland) are reproduced from Ref. [15]. Data shown as blue diamonds (UBC Nano-g) are reproduced from Ref. [22]. Data shown as blue squares (QNS block-Seoul), triangles (IBM-ETH Zurich), and stars (MPI-Stuttgart) are reproduced from Ref. [2].

Narrow band spectra of the floor velocity in *X*, *Y*, and *Z* directions are shown in **Figure 3**, comparing measurements on the floating block with those on the lab floor before construction, and after isolation of the foundation base plate. We find that the floor vibration levels in the cradle are attenuated by a factor of ~4 in both the vertical and horizontal directions across the entire investigated frequency range. However, the main improvement in vibration levels is achieved on the floating block, after active airspring isolation. The active isolators effectively reduce residual low-frequency vibration impact by up to two orders of with the single-axis magnitude above ~3 Hz when compared to the foundation base plate. Below 3 Hz, the inherent noise of the isolation system becomes visible, since the floor vibrations of the laboratory have reached very small levels structurally. As a result, we find that the active control of the air springs is only partially able to suppress the low frequencies resonance. However, given that the resonance frequencies of both STM table (passive stage) and the STM itself are >3 Hz, we observe minimal impact of this resonant amplification as reflected in measurement of the STM's junction noise (see **Figure 5**).

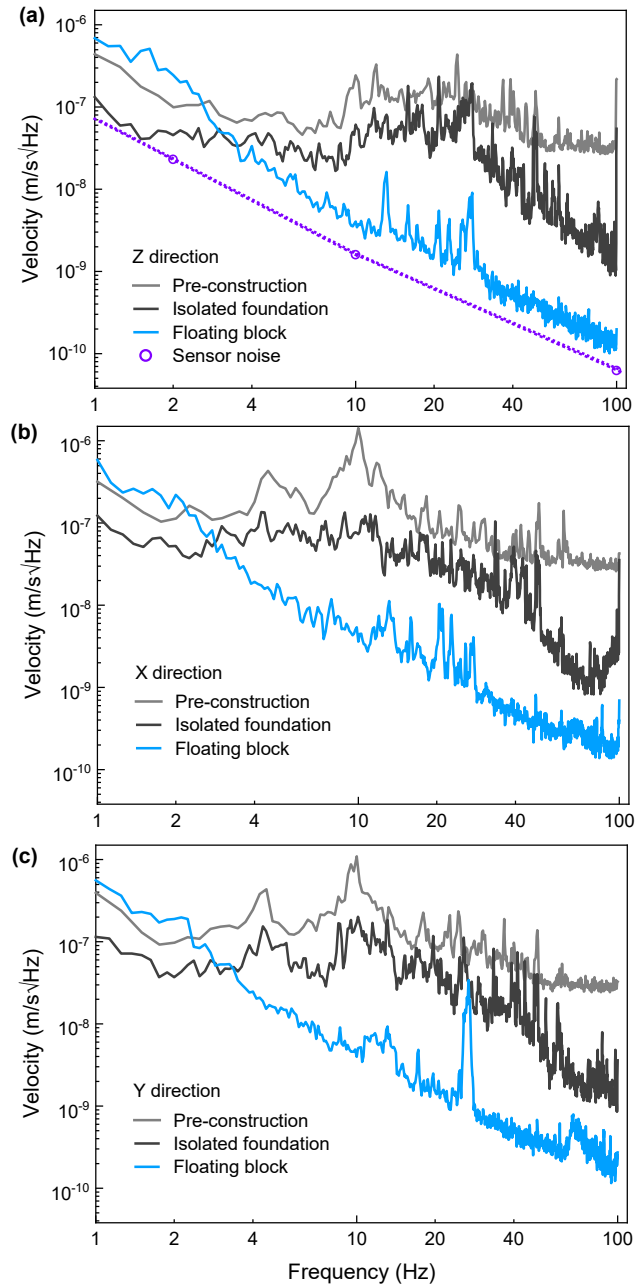

**Figure 3. Narrow band spectra of the floor velocity. (a-c)** Comparison of the spectral density in peak-hold, measured on the floor before construction (light gray), on the isolated foundation base plate/cradle (black), and on the floating block (blue), respectively, in both vertical **(a)** and horizontal **(b, c)** directions. The dashed line in **(a)** indicates the seismic sensor noise level. The vibration data measured before the construction (light gray) were recorded by a 3-axis seismic sensor (Wilcoxon 731-207) with lower sensitivity compared Wilcoxon 731A used for all other seismic measurement in this work, and are seen to saturate at $(3 \times 10^{-8})$ m/s√Hz for frequencies above 50 Hz.

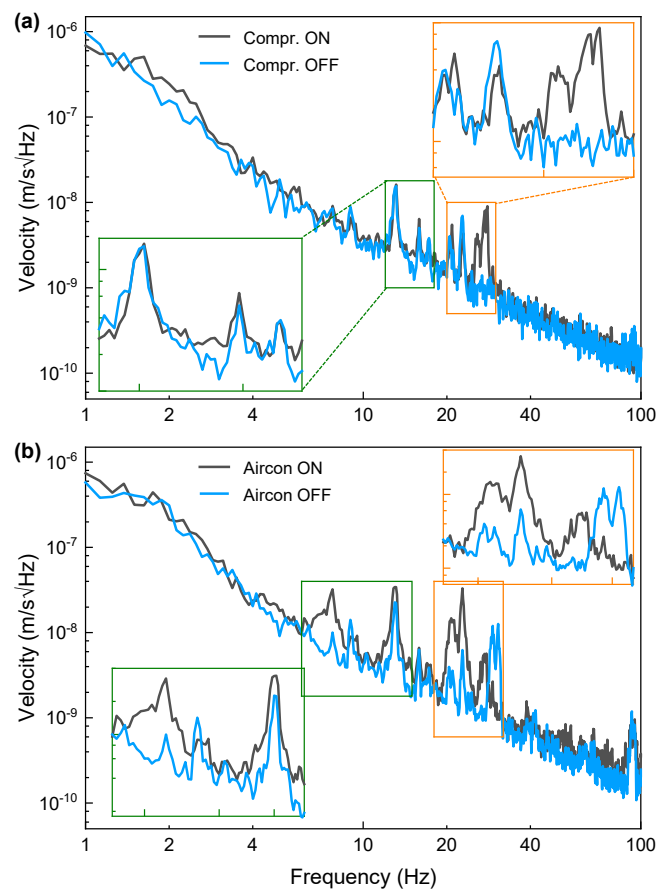

**Figure 4. Impact of air compressor and lab air-conditioning. (a)** Comparison of the spectral density of the vertical floor velocity (peak-hold), measured on the floating slab with the air compressor on and off and with air-conditioning airflow turned off. The compressed air (initial pressure: 10 bar) stored in the buffer tank can maintain the pressure above the working pressure (5 bar) of the air springs for a couple of minutes after the air compressor is turned off, allowing us to complete the vibration measurement. **(b)** Comparison of the spectral density of the floor velocity (peak-hold), measured on the floating slab with the air- conditioning airflow switched

on and off. During this measurement the air compressor was running. The Inserts in **(a)** and **(b)** show zoomed-in spectra to highlight resonance peaks.

We note the presence of a few resonances in the range of 10 Hz to 30 Hz on the floating slab (**Figure 3**). Vibration impact within this frequency range [23] is usually attributed to airborne acoustic noise or may be due to pressure fluctuation in the compressed air to the air springs. To uncover the source for such impact, in **Figure 4a** we compare vibration spectra measured with the air compressor running and with it switched off. We find that the resonance around 27 Hz is fully eliminated once the air compressor is off, while the remainder of the spectra is of comparable magnitude. This suggests that air compressor might generate pressure fluctuations in the compressed air system, coupling to the block To evaluate the impact of air conditioning induced airflow turbulence or pressure fluctuations, in **Figure 4b** we show spectra with airflow switched on and off. The overall vibration levels remain of comparable magnitude; however, we find mild impact at 8 Hz, 13 Hz, 21 Hz, and 23 Hz.

*3.2 STM performance benchmark*

The ultimate performance benchmark of an ULV laboratory for DR-cooled STM are noise and vibration levels as detected by the STM's tunnelling junction after all layers of isolation. Spectra of the STM tunnelling current (feedback loop open) and piezo extension *Z* (feedback loop closed) are summarized in **Figure 5a-b**. Typical feedback parameters with proportional gain of 1~3 pm and time-constant of 100 μs were used. A Specs Nanonis RC5e control system was used to control the STM and a low noise Femto DLPCA-200 preamplifier with gain of $10^9$ V/A and bandwidth of ~1 kHz was used to record the tunnelling current. All data were recorded with a Pt-Ir tip on single-crystal Au(111).

Our vibration-isolation setup allows four different configurations of isolation stages depending on whether the active (floating slab) or passive (STM table) stages are activated.

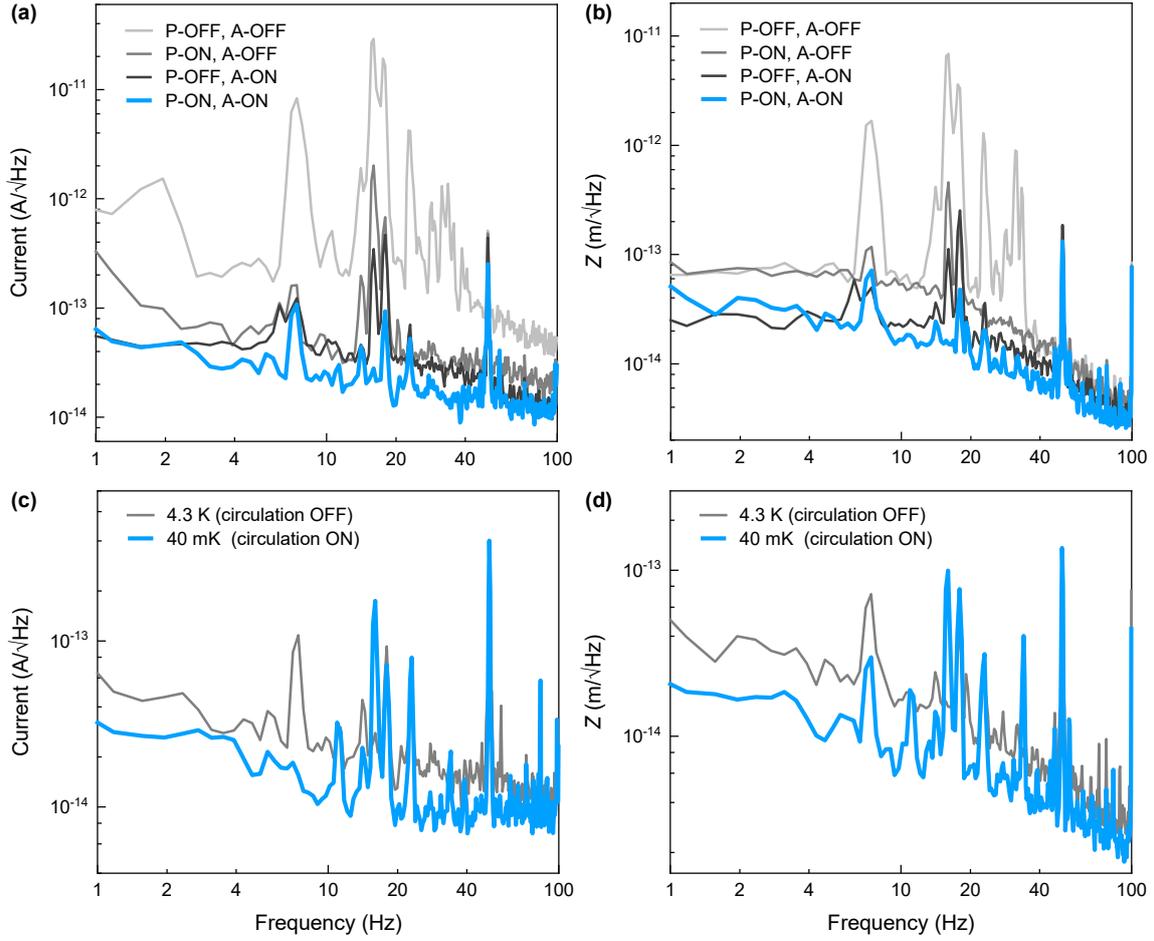

**Figure 5. Effect of vibration isolation on the STM junction noise. (a-b)** Spectral density of the tunnelling current **(a)** and the $Z$ height **(b)** at 4.3 K for different combinations of the passive and active stages. The $Z$ spectra were recorded on clean Au(111) with the feedback loop closed (proportional gain of 1~3 pm, and a time constant of 100 μs) and with tunnelling conditions: $V_{bias}$ = 0.2 V, $I$ = 100 pA. The current spectra were recorded with the feedback loop switched off and the tip in tunnelling range. P-ON/OFF: passive isolator (STM table) is floating/resting; A-ON/OFF: active isolator (concrete block) is floating/resting. **(c-d)** Spectral density of the tunnelling current **(c)** and $Z$ **(d)** at 4.3 K (40 mK), with both STM table and block floating, and with DR circulation off (on).

A comparison is shown in **Figure 5a-b**, highlighting the respective efficiency in vibration isolation. As seen from the data, both stages combined are able to reduce peak noise level by factors of 240 (80) in current ($Z$), respectively, while further reducing the low-frequency noise floor by factor of 5 (2). Remaining peaks at 8 Hz and 18 Hz might reflect the natural frequencies of the STM which fall into this range. Vibration impact from the lab floor and surrounding building at 13 Hz and 21 Hz also seen in **Figure 3** are effectively attenuated. We note that all spectra show a sharp peak at 50Hz almost independent of vibration isolation, which reflects the electrical power line cycle noise. The lowest noise levels achieved with both stages activated are of the order ~10 fA/√Hz at <100 Hz. Even the elevated noise levels at 50 Hz, below 200 fA/√Hz (100 fm/√Hz) in current (*Z*), are comparable or better than published results for state-of-the-art DR-cooled STMs [14,15,24], and may be optimized in the future. Attenuation of the junction noise by one order of magnitude or more at all frequencies, detected in both current and *Z* reflects that lab design and isolation stages are very effective in improving the overall STM performance, even for a commercial STM of mature design, such as the USM1600.

The vibration impact of the DR circulation pumps on the STM's junction noise is summarized in **Figure 5c-d**, where we compare data measured at 4.3 K (DR circulation off) with that measured at base temperature (40 mK) with the DR circulation on. Surprisingly, both the current and the *Z* spectra show a lower noise floor at base temperature despite operation of the circulation pumps. This result is reproducible across multiple tip-sample combinations and illustrates very effective vibration isolation of our gas handling system via the T-junction and soft bellows. A single peak at 16 Hz, appearing in both the current and *Z* noise, not observed at 4.3 K, confirms weak transmission of pump noise to the STM at this frequency.

Basic STM performance at the base temperature of the DR is summarized in **Figure 6**, where we quantify the junction temperature, a key limiting factor to the STM's spectral resolution. In an ideal noise-free environment, an STM's spectral resolution $\Delta E = 3.5\, k_B T$ is only limited by temperature. Here $k_B$ is the Boltzmann's constant, and the factor of 3.5 reflects thermal broadening of the sample's local density of states (LDOS) as quantified by the width of the first derivation of the Fermi-Dirac distribution. However, external sources of electronic or mecha-nical noise may give rise to a further reduction in spectral resolution due to junction heating or a reduced signal-to-noise level. **Figure 6a** shows the measured LDOS of a superconducting polycrystalline aluminum (Al) sample. The solid blue line shows a fit to Maki theory [12,24,28,29] which, in the absence of an external magnetic field, expresses the superconducting DOS as

$$\rho(E) = \text{Re}\left(\frac{u}{\sqrt{u^2-1}}\right), \qquad (1)$$

where $u$ is defined by the nonlinear equation,

$$u = \frac{E}{\Delta} + \zeta \frac{u}{\sqrt{1-u^2}} \qquad (2)$$

with superconducting gap $\Delta$ and pair-breaking parameter $\zeta$. Convolving equation (1) with the derivative of the Fermi-Dirac distribution allows us to extract a gap of $\Delta = (179.7 \pm 0.2)$ µeV and pair-breaking parameter $\zeta = (0.015 \pm 0.009)$, consistent with previous reports [12,29]. We further extract an effective temperature $T_{\text{eff}} = (144.3 \pm 0.5)$ mK, with a corresponding energy resolution of $3.5 k_B T = 43$ µeV, comparable or better than published data for custom-built [12–15,24,29–31] or commercial [32] DR-cooled mK- STMs. We have separately fitted the data to a BCS-Dynes theory [33] (not shown), which results in slightly reduced fit quality but a similar estimate for the junction temperature $T_{\text{eff}} = (160.2 \pm 0.1)$ mK.

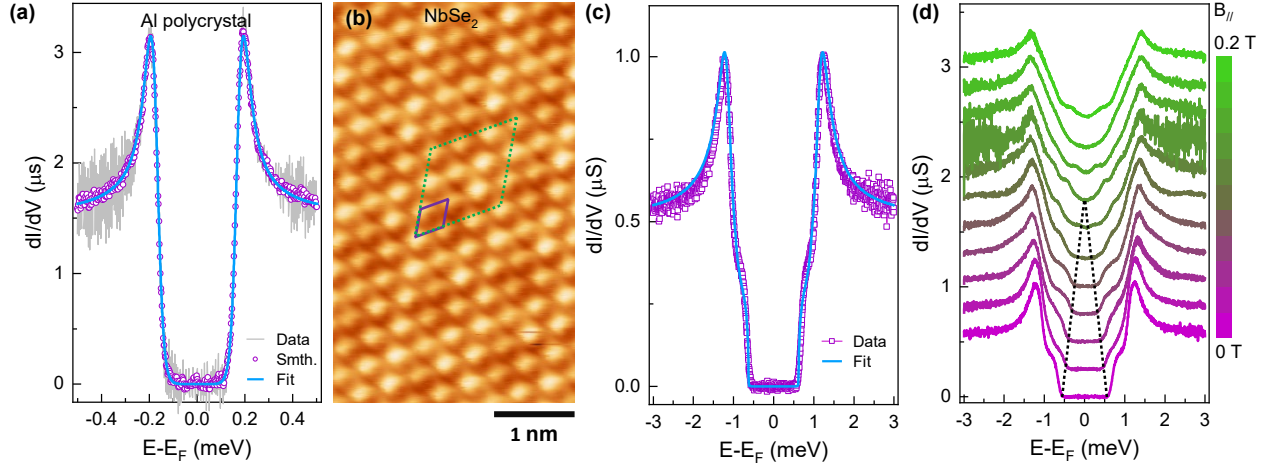

**Figure 6. STM performance at mK temperatures.** (**a**) STS spectrum of superconducting polycrystalline aluminum (Al), used to extract the junction temperature. The solid line shows a Maki fit, from which we obtain $T = (144.3 \pm 0.5)$ mK. The raw and smoothed data are shown as gray line and purple markers, respectively. (**b**) Atomic-resolution topography image of $NbSe_2$. The purple solid and green dashed rhombus highlight the unit cells of Se sublattice and the 3×3 charge density wave order, respectively. The image was taken under constant current mode. (**c**) Superconducting local density of states (LDOS) of $NbSe_2$. The solid line shows a fit based on the McMillan equations [25–27] with the temperature fixed to 150 mK. (**d**) Evolution of superconducting DOS of $NbSe_2$ under in-plane magnetic fields up to 0.2 T. The spectra are offset vertically for clarity. All data were taken at the base temperature ~30 mK with both the STM table and concrete block afloat, as well as all utilities running including air compressor and air conditioning (standard lab operation). The differential conductance (d$I$/d$V$) spectra in (**a**), (**c**) and (**d**) were taken using standard lock-in techniques with an AC modulation frequency of 973 Hz and a modulation amplitude $V_{mod}$ for 3 μV(**a**), 10 μV (**c**, **d**), with a integration time of 10 ms (**a**) and 5 ms (**c**, **d**), respectively. A small modulation amplitude was used in the measurement to minize the lock-in induced instrumental broadening to the spectra. The purple markers in (**a**) were smoothed with a 5-point-window weighted average to simulate the broadening of a 10 μeV lock-in amplitude as used to acquire the data in (**c**, **d**). The tunnelling setpoint conditions were (**a**) $V_{bias}$ = -1 mV, $I$ = 2 nA, (**b**) $V_{bias}$ = 10 mV, $I$ = 1 nA, and (**c**, **d**) $V_{bias}$ = -3 mV, $I$ = 1.5 nA, respectively.

To further demonstrate spectral resolution of our mK-STM and operation of the magnetic coil, we measured the superconducting DOS of the layered type-II superconductor NbSe$_2$ [34]. An atomic-resolution image of the NbSe$_2$ surface is shown in **Figure 6b**. The NbSe$_2$ Fermi surface is composed of three bands [35–37], out of which the two Nb bands are believed to carry superconductivity. As a result of the two-band nature, two distinct superconducting energy gaps, $\Delta_L$ and $\Delta_S$, are expected at $T \ll T_C$ [26,38,39]. This is clearly reflected in the measured superconducting LDOS of NbSe$_2$ as shown in **Figure 6c**. From the fits to the McMillan equations [25–27], we extract gap parameters $\Delta_S = (0.38 \pm 0.02)$ meV, and $\Delta_L = (1.14 \pm 0.01)$ meV, in very good agreement with previous work [26,27,39]. Noise performance under application of an in-plane magnetic field is demonstrated **Figure 6d**, which shows the evolution of the superconducting DOS between $B = 0$ T and $B = 0.2$ T, where the smaller of the two superconducting gaps is seen to close at $B \sim 0.1$ T.

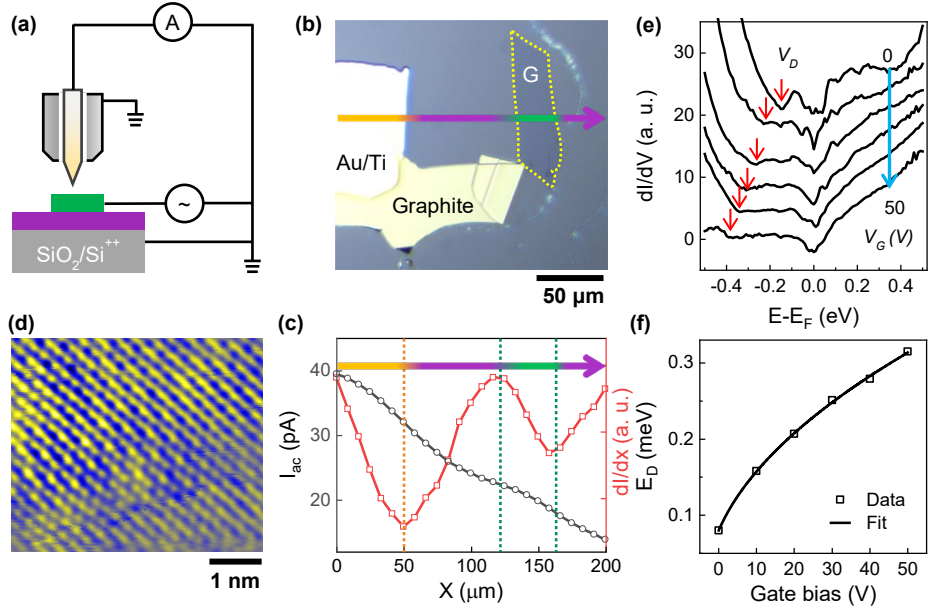

**Figure 7. Tip approach onto a micron-sized sample via capacitance-mediated detection.** **(a)** Schematic circuit for a capacitance-mediated tip approach. **(b)** Optical image of an exfoliated graphene sample with Au/Ti contact. The graphene flake outlines are highlighted by the yellow dashed lines. **(c)** AC current profile (black) and its derivative (red) measured while the STM tip is scanning along the direction as indicated in **(b)**. The AC current was recorded by a lock-in amplifier with an amplitude of 5 V at 1.726 kHz. The derivative of the AC current was numerically obtained. **(d)** Atomic-resolution topographic image of the exfoliated graphene crystal. **(e)** Differential conductance spectra under application of a gate bias ($V_G$) from 0 V to 50 V. The red arrows indicate the Dirac point ($V_D$) of graphene. d$I$/d$V$ spectra were taken using standard lock-in techniques with an AC modulation of 10 mV at 923.5 Hz using an integration time of 100 ms. **(f)** $E_D$ vs gate bias. The data are extracted from (e) by $E_D = V_D - \hbar\omega$ [40], where $\hbar\omega$ is the energy of phonons involved in the inelastic tunnelling process. The fitting is based on the formula $E_D = \hbar v_F\sqrt{\pi\alpha|V_G-V_0|}$ [40], from which we extract the charge neutral point $V_0 = -3.5$ V, indicating slight $n$-doping in graphene. The tunneling setpoint conditions were (d) $V_{bias}$ = 40 mV, $I$ = 1 nA, and (e) $V_{bias}$ = 0.5 V, $I$ = 1 nA.

The capability to approach a tunnelling tip onto a micron-sized sample significantly enriches the scope of STM studies, given that a plethora of exotic low-temperature phenomena [41,42] have been predicted for 2D crystals and their heterostructures. More so, such capability would allow multiple electrodes to be integrated to contact and control a micron-sized electrical device, e.g., by application of gate voltages [40,43–49]. As access with long focal optical microscopes [40] is not easily possible in the DR environment, here, we employ a capacitance-mediated approach [50] for landing the STM tip onto a micron-sized sample. **Figure 7** summarizes our demonstration of capacitance-mediated tip approach onto a basic back-gated graphene transistor on a $SiO_2$/Si substrate (**Figure 7b**). As shown in **Figure 7a**, an AC voltage is applied to the graphene sample, resulting in an AC current passing through the tip that can be detected by a lock-in amplifier. As the tip is not in tunnelling range, this current depends on the imaginary part of the STM's junction impedance, dominated by the capacitance between tip and sample. This capacitance is composed of two dominant parts, $C = C_{ST}(x, y) + C_{SH}$, i.e., the sum of the capacitance between tip and sample ($C_{ST}$) and that between tip holder and sample ($C_{SH}$). The first part strongly depends on the tip location with respect to the sample when the tip is very close (<1/3 of the sample's lateral size [50]), allowing to locate metallic features. On the other hand, $C_{SH}$ is somewhat independent of the tip location since the tip holder is usually much larger than the tip (the diameter of the tip holder is ~5 mm, compared to 0.3~0.5 mm for the tip), which thus results in an increase of background noise. To maximize the capacitance signal from the tip, we therefore employ a customized tip holder with a grounded shield (**Figure 7a**) developed by UNISOKU, such that the capacitance between sample and tip remains dominant. A measurable change in ac current is observed when moving the tip across the conducting sample edges as illustrated in **Figure 7b-c**, more clearly visible in d*I*/d*x*, allowing to identify the sample location by comparison with the corresponding optical image of the device. **Figure 7d** shows an atomic-resolution topographic image of the graphene lattice after successfully landing the tip via the capacitance-mediated approach. As shown in **Figure**

**7f-e**, a gate bias applied to the doped Si back-gate, allows to tune the carrier density in the graphene, as indicated by a shift of Dirac point with gate bias.

## 4. Conclusion

In summary, we have reported the design and performance benchmarking of an ultra-low vibration (ULV) laboratory hosting a commercially available UNISOKU USM1600 40-mK STM. We have adopted a "room-within-a-room" concept for the laboratory, placed on a separate foundation from the surrounding building. Central part of our vibration isolation is a T-shaped concrete mass block suspended by actively controlled pneumatic springs. Vibration measurement reveals that the laboratory achieves extremely low vibration levels, meeting the VC-M vibration standard for frequency higher than 3 Hz reached in only a limited number of lab world-wide. Measurement of the STM's junction noise reflects the effective vibration isolation, even with the laboratory's services running including all dilution refrigerator pumps and air conditioning, allowing for an energy resolution as low as 43 μeV (144 mK). The successful design of an ultra-low vibration laboratory for DR-cooled STM promises for high-precision investigation and control of quantum matter at atomic length scales.

**Acknowledgements**

This work is supported by National Research Foundation (NRF) Singapore, under the Competitive Research Programme "Towards On-Chip Topological Quantum Devices" (NRF-CRP21-2018-0001), with partial support from a Singapore Ministry of Education (MOE) Academic Research Fund Tier 3 grant (MOE2018-T3-1-002). B.W. acknowledges a Singapore National Research Foundation (NRF) Fellowship (NRF-NRFF2017-11).